\documentclass[twocolumn,twoside,slac]{revtex4}
\usepackage{graphicx}
\usepackage{fancyhdr}
\begin{document}

\title{ The Data Acquisition System Based on PMC Bus}

%

\author{
Y. Igarashi, H. Fujii, T. Higuchi, M. Ikeno, E. Inoue, R. Itoh, H. Kodama,
T. Murakami, M. Nakao, K. Nakayoshi, M. Saitoh, S. Shimazaki,
S. Y. Suzuki, M. Tanaka, K. Tauchi, M. Yamauchi, Y. Yasu
}
\affiliation{KEK, Oho 1-1, Tsukuba, Ibaraki 305-0801, Japan}
\author{Y. Nagasaka}
\affiliation{Hiroshima Institute of Technology,
Miake 2-1-1 Saeki Hiroshima 731-5193, Japan}
\author{G. Varner}
\affiliation{University of Hawaii, 2505 Correa Rd., Honolulu, HI 96822, USA}
\author{T. Katayama, K. Watanabe}
\affiliation{Densan Co. Ltd.,
Takaido 1-25-16, Suginami, Tokyo 168-0074, Japan}
\author{M. Ishizuka, S. Onozawa, and C. J. Li}
\affiliation{Designtech Co. Ltd.,
Higashisumiyoshi 18-9, Tokorozawa, Saitama 359-1124, Japan.}

\begin{abstract}
        High energy physics experiments in KEK/Japan rush into over kHz
trigger stage.  Thus, we need a successor of the data acquisition(DAQ)
system that replaces the CAMAC or FASTBUS systems.  To meet these needs,
we have developed a widely usable high-density DAQ system which includes
a crate, base-board, daughter cards for front-end A/D or T/D conversion, and
back-end communication base-board for data transfer and timing control.
The size of the crate is for the 9U Euro-cards with the standard VME32 bus
and extension connectors for power supply.  The base-board comprises of
a local bus with the sequencer connected to the front-end daughter cards
via event buffering FIFOs, and the standard PMC (PCI mezzanine card)
bus to be set a PMC processor unit to reduce data size from the front-end
daughter cards.  A data transfer module, which is connected to the
event building system, and a trigger control unit, which communicates
with the central timing controller are installed on the back-end
communication card connected to the rear end of the base-board.
We describe the design of this DAQ system and evaluate the performance of it.
\end{abstract}

\maketitle

\thispagestyle{fancy}


\section{Introduction}

The data acquisition (DAQ) system of the next generation
experiments proposed in KEK/Japan must operate with trigger rates
significantly higher than the currently on-going experiments.
Numbers of experiments are planned at a high intensity 50 GeV proton
accelerator project (J-PARC\cite{j-parc}).
The KEKB B-factory also plans to upgrade to a luminosity ten times higher
 ${\cal L} = 10^{35}{\rm cm^{-2}s^{-1}}$(Super KEKB\cite{superb}).
The trigger rate of these experiments are estimated as 1 kHz $\sim$ 10 kHz
for the J-PARC experiment and $\sim$10 kHz for upgraded the KEKB experiment.
In the above ranges, conventional DAQ systems based on CAMAC,
Fastbus, or VME are not appropriate for our needs:
CAMAC does not allow high density channels because of its small board size;
Fastbus is commercially dying out;
and VME does not supply adequate power for analog signals.
Because there is no suitable (de-facto) standard system capable of working
 at such a trigger rate,
we developed a DAQ platform which may have a wide range of uses.
The first application is the DAQ system for experiments
at J-PARC and the Super KEKB experiment.

\section{Design overview}

We design the DAQ platform with the following features:
\begin{itemize}
\item Front-end Buffering: \\
 Signals from detectors are buffered in the front-end while awaiting
 a trigger decision and the timing signals for data transfer are uniformly
 distributed to the back-end.
 This is effective at reducing the dead-time.
\item Trigger/busy handshake: \\
 The Trigger/busy handshake works at a trigger rate of up to 10 kHz.
\item Front-end data reduction: \\
 Large amounts of data generated by Flash ADC(FADC),
 can be reduced on the board.
\item Data transporting by data link: \\
 Data are transferred downstream to the DAQ system via  a standard data link
such as a Fast Ethernet.
\item Modular system: \\
The system is composed of modules for maximum flexibility,
allowing choice of digitization scheme, processor and data link. 
Being modular also makes it easier to upgrade,
for example, the processor or data links.
\item Wide scalability: \\
  The same DAQ system can be used not just for small-scale tests
but also for large-scale experiments such as the Super KEKB experiment
\item Cost effectiveness
\end{itemize}

We developed a 9U-size power crate with a
9U-size main board and a 6U-size rear board as the base elements.
The power crate has a standard VME32 bus and an extension connector 
for the power supply.
The main board is equipped with a PCI mezzanine card (PMC) bus so that the
fastest processor available can be used.
The main board has three PMC slots and four front-end daughter card slots.
Each of the PMC slots and each of the daughter card slots have a FIFO buffer
for efficient data handling.
The PMC buses run from the main board to the rear board.
The rear board has two PMC slots and a special slot for the trigger
distribution card.
The front-end function, such as FADC or pipeline TDC,
can be installed as front-end daughter cards similar in size to PMC cards.
The event signals digitized by the front-end cards are buffered in FIFOs,
transferred to the processor via the PMC bus, reduced by the processor,
and then sent to the event-builder system through the data transfer interface
via the PMC bus.

\begin{figure}
\centering
\includegraphics[width=65mm]{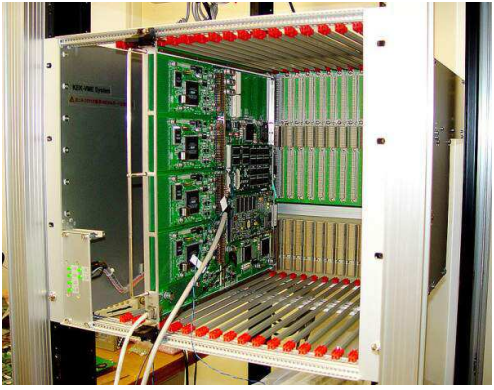}
\caption{KEK-VME crate with a read-out module}
\label{cratepicture}
\end{figure}

The trigger information is received by the trigger card
on the rear board and distributed to each front-end card.
Figure~\ref{cratepicture} shows the crate and
a read-out module used for the test.

Details of the DAQ platform components are described
in the following sections.

\section{KEK-VME crate, an extension of standard VME}

 We chose a 9U Euro card with a VME32 bus as the base of our system
because these are widely used and are cost effective.
However, the default VME system cannot provide sufficient current
needed for digitization analog devices.
Moreover, the standard VME power supply is too noisy
for delicate handling of the analog signals.
Therefore, we added a dedicated connector
as J0 for the power supply, to avoid any concern and to simplify
the power treatment.
Supplied voltages via the J0 connector are -5.0 V and $\pm$3.3 V.
Maximum currents delivered by J0 connector are 100A for -5.0 V, 320 A
for +3.3 V and 200 A for -3.3 V.
KEK-VME J0 connector is used as same as VME64x J0 connector.
The definition of pins are shown in Table~\ref{j0pin}.
Eight pairs of the differential-signal bus line for the 100-Mbps LVDS
are assigned as S1$\pm$, S2$\pm$, S3$\pm$, S4$\pm$,
S5$\pm$, S6$\pm$, S7$\pm$ and S8$\pm$ in the J0.
Its line impedance of the differential-signal bus line is 100 Ohm.

\begin{table}
\begin{center}
\caption{KEK-VME J0 connector pin-out definition}
\label{j0pin}
\begin{tabular}{|r|ccccccc|}
\hline
Pos.	& z	& a	& b	& c	& d	& e	& f \\
\hline
1	& GND	& GND	& GND	& GND	& GND	& GND	& GND \\
2	& GND	& GND	& GND	& GND	& GND	& GND	& GND \\
3	& GND	& GND	& GND	& GND	& GND	& GND	& GND \\
4	& GND	& +3.3V	& +3.3V	& +3.3V	& +3.3V	& +3.3V	& GND \\
5	& GND	& +3.3V	& +3.3V	& +3.3V	& +3.3V	& +3.3V	& GND \\
6	& GND	& +3.3V	& +3.3V	& +3.3V	& +3.3V	& +3.3V	& GND \\
7	& GND	& +3.3V	& +3.3V	& GND	& GND	& GND	& GND \\
8	& GND	& GND	& GND	& GND	& GND	& GND	& GND \\
9	& GND	& GND	& GND	& GND	& GND	& GND	& GND \\
10	& GND	& GND	& GND	& GND	& -3.3V	& -3.3V	& GND \\
11	& GND	& -3.3V	& -3.3V	& -3.3V	& -3.3V	& -3.3V	& GND \\
12	& GND	& -3.3V	& -3.3V	& -3.3V	& -3.3V	& -3.3V	& GND \\
13	& GND	& GND	& GND	& GND	& GND	& GND	& GND \\
14	& GND	& -5V	& -5V	& -5V	& -5V	& -5V	& GND \\
15	& GND	& GND	& GND	& GND	& GND	& GND	& GND \\
16	& GND	& S1+	& S1-	& GND	& S2+	& S2-	& GND \\
17	& GND	& S3+	& S3-	& GND	& S4+	& S4-	& GND \\
18	& GND	& S5+	& S5-	& GND	& S6+	& S6-	& GND \\
19	& GND	& S7+	& S7-	& GND	& S8+	& S8-	& GND \\
\hline
\end{tabular}
\end{center}
\end{table}


\begin{figure}
\centering
\includegraphics[width=65mm]{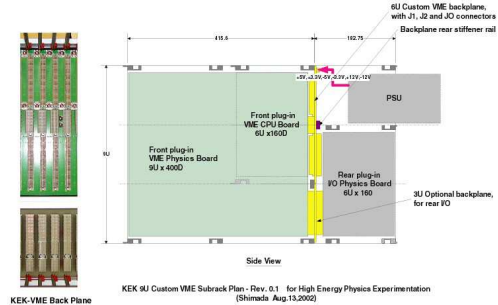}
\caption{KEK-VME crate and it's part of back-plane}
\label{kekvme}
\end{figure}

Moreover, we are also developing a low noise and cost-effective power supply.

\section{Read-out module}

FADC and pipeline TDC generate large amounts of data.
To reduce the data at an early stage, a pipeline buffer and a processor
were installed on the board.

The read-out module meets the following requirements:

\begin{itemize}
\item A upgradeable processor:
 because processors are evolving rapidly.
\item A changeable data transfer interface:
 because this technology is also evolving rapidly and experimental
 setups call for different requirements.
\end{itemize}

To satisfy these requirements, we have developed a new DAQ system
equipped with a PMC for the digital part of
the read-out module.

The system consists of four components:
two types of base boards, 
a front-end mezzanine card, and a PMC .
One of the base boards is named COPPER
(COmmon Pipelined Platform for Electronics Readout).
This has four slots for the front-end mezzanine cards on half of the
front side, a PMC slot for the processor, and two universal PMC slots
on half of the back side.
The other is a rear board named SPIGOT
(Sequencing Pipeline Interface with Global Oversight and Timing),
which has two universal PMC slots and a slot for a trigger module.
A front-end mezzanine card named FINESSE
(Front-end INstrumentation Entity for Sub-detector Specific Electronics)
is a daughter card with A/D converting function. 
Various types of FINESSE have been made, such as multiple TDC and FADC.

\begin{figure}
\centering
\includegraphics[width=65mm]{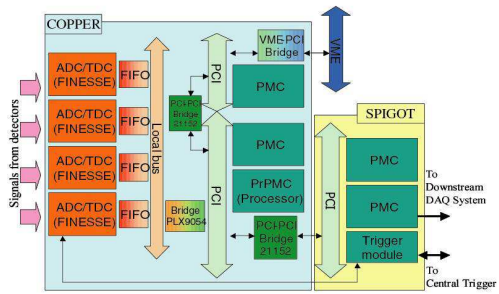}
\caption{A schematic view of read-out module}
\label{moduledesign}
\end{figure}

\subsection{PMC(PCI Mezzanine Card)}

The PMC, defined by IEEE1386.1\cite{pmc},
separates the processor unit and the read-out module.
Its functions and electrical characteristics are the same as a PCI,
but it measures only 74 mm x 149 mm.
It can be used with VME, VME64, VME64x, Compact PCI, etc.
Several modules are available at reasonable prices, as the technology
involved is the same as that for PCIs used in common PCs.

\subsection{Base board}

\begin{figure}
\centering
\includegraphics[width=65mm]{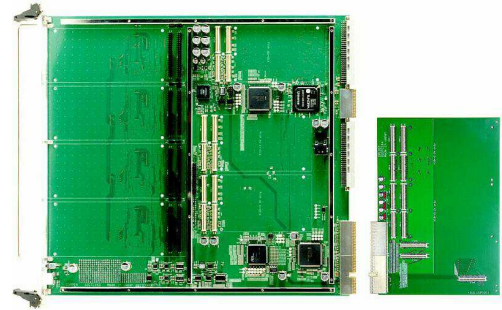}
\caption{The COPPER and SPIGOT base boards}
\label{baseboard}
\end{figure}

One of the base boards is a 9U-size main base board designated
COPPER (COmmon Pipelined Platform for Electronics Readout).
The other is a 6U-size rear base board designated SPIGOT
(Sequencing Pipeline Interface with Global Oversight and Timing).
The boards are shown in Figure~\ref{baseboard}.

The front half of the COPPER is for signal digitizing and the back half
is for data processing. The front has four slots
for front-end mezzanine cards, which are described below. 
The back has a PMC slot for the processor and two universal PMC slots.
A bus bridge, using PLX technology incorporated in the PCI 9054\cite{pci9054},
connects the front local bus and the back PMC bus.

Eight 256KB FIFOs receive data from the mezzanine cards
which are passed to the memory on the processor board
by a sequencer on the local bus and the PLX PCI-9054.

The SPIGOT is at the back and is connected to the COPPER via a PCI bus
and trigger information lines.
The SPIGOT has a trigger information module and a PMC card for
a data transfer interface such as Ethernet or a serial link(IEEE1394, etc).

\subsection{Processor PMC}

As the processor PMC card, we use a Radisys EPC-6315\cite{epc6315}.\\
The specification of the processor are:.
\begin{itemize}
\item 800 MHz Pentium III-M Processor
\item up to 512 MB PC-133 compliant SD-RAM with ECC
\item 10/100 BaseT Ethernet port
\item On-board CompactFlash socket
\item 32-bit 33/66 MHz PCI bus interface
\end{itemize}

We chose PC architecture for front-end data processing,
because PC architecture has the following merits:

\begin{itemize}
\item It is familiar to many people because it has the same architecture
 as the common PC.
\item PCs can run operating systems rich in function.
 Therefore we can develop applications easily and drive many complex devices
 (such as Ethernet TCP/IP).
\item PC are common throughout the world, and fast,
 yet cheap, processors are readily available.
\item A large amount of knowledge is openly available.
\end{itemize}

\begin{figure}
\centering
\includegraphics[width=65mm]{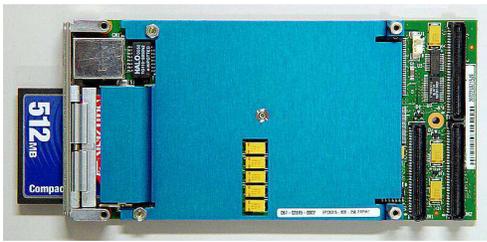}
\caption{PC architecture Processor PMC board Radisys EPC-6315}
\label{epc6315}
\end{figure}

Processor PMC card EPC-6315 shown in Figure~\ref{epc6315}.

We use Linux 2.4 on a on-board processor.
Linux is a most popular development environments in
high energy physics experiments.

\subsection{FINESSE, Front-end daughter card}

\begin{figure}
\centering
\includegraphics[width=65mm]{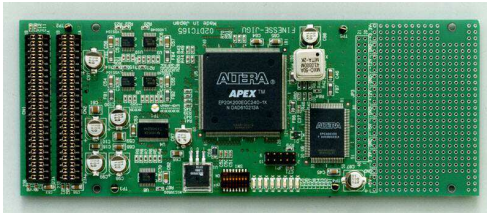}

\caption{FINESSE-jig, a front-end daughter card for testing system}
\label{finessejig}
\end{figure}

The FINESSE
(Front-end INstrumentation Entity for Sub-detector Specific Electronics)
front-end daughter card for testing the system was developed first,
instead of a real FADC or pipeline TDC.
It is called a FINESSE-jig and has FIFO logic and trigger/busy handshaking
logic integrated on it.
It generates the simulated data, and
stores it into local FIFO and exports it to the other FIFO
on the base board upon receiving a trigger.
FINESSE-jig is shown in Figure~\ref{finessejig}.

FINESSE is being further developed.
A recently developed FINESSE is the FINESSE AMT2 which is a TMC
(time memory cell)\cite{tmc} based pipeline TDC card, has an AMT2 chip.
The AMT2(ATLAS Muon TDC 2)\cite{amt2} chip is the latest TMC chip
developed for ATLAS muon chamber. 
Its specifications are shown in Table~\ref{finesseamtspec}

\begin{table}
\begin{center}
\caption{Specifications of FINESSE AMT2}
\label{finesseamtspec}
\begin{tabular}{|l|l|}
\hline
TDC & AMT2 \\
Input &  24 ch LVDS \\
Time resolution & 0.78 ns/bit (at using 40 MHz clock) \\
Trigger buffer depth & 8 words \\
\hline
\end{tabular}
\end{center}
\end{table}

\begin{figure}
\centering
\includegraphics[width=65mm]{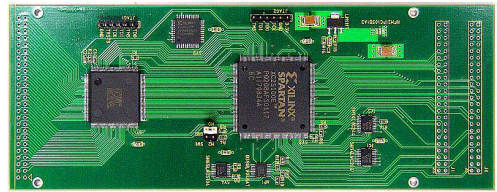}
\caption{FINESSE AMT2, a pipeline TDC front-end daughter card}
\label{finesseamt2}
\end{figure}

FINESSE AMT2 is shown in Figure~\ref{finesseamt2}

Another recently developed FINESSE is the FINESSE FADC which uses
for four FADC chips, Analog Devices incorporated AD9235-20.
Its specifications are shown in Table~\ref{finessefadcspec}

\begin{table}
\begin{center}
\caption{Specifications of FINESSE FADC}
\label{finessefadcspec}
\begin{tabular}{|l|l|}
\hline
ADC & Analog Devices AD9235-20 \\
Number of channel &  8 ch \\
Resolution & 12 bit \\
Maximum sampling clock & 20 MHz\\
\hline
\end{tabular}
\end{center}
\end{table}

\section{Performance and stability}

We have tested the on-board PCI bus and local bus system for 
data transfer speed, stability over long periods of continuous use, and 
thermal stability.

 We measured the speed of data transfer from the FINESSE
to the processor main memory in DMA mode, using the FINESSE-
jig.  The measured transfer speed without overhead was
125 MB/sec.  This result shows that it works with 95\% of
the 32bit 33MHz PCI bus performance.  We also measured
the transfer speed including the overhead of the sequencer
on the local bus.  The speed was 80 MB/sec when the
event size was 200 bytes.

To test for module reliability,
we ran the system for 77 hours continuously.
To check for error, all the transferred data were compared
with data generated by FINESSE-jig.

\begin{figure}
\centering
\includegraphics[width=65mm]{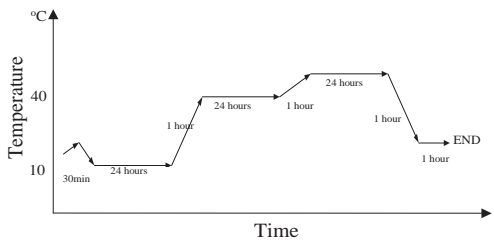}
\caption{Temperature variation in the oven during the stability test}
\label{tempgraph}
\end{figure}

To test the thermal stability, we placed the read-out module
in a thermostatted oven and changed the temperature as
shown in Figure~\ref{tempgraph}.

\begin{figure}
\centering
\includegraphics[width=65mm]{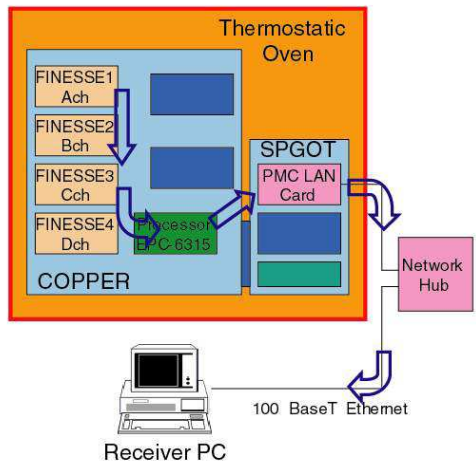}
\caption{Setup for stability tests}
\label{testsetup}
\end{figure}

Data generated by FINESSE-jig were read by FINESSE FIFO,
transferred by the PLX PCI-9054 DMA via a local bus and
a PCI bus to the processor main memory,
and then sent using a LAN card via a 100 BaseT network to
another computer outside the oven.
In order to simulate actual use, we made the setup
as shown in Figure~\ref{testsetup}.
Data generated by FINESSE-jig were
stored into FINESSE FIFO, transferred by the PLX PCI-9054 DMA via a
local bus and a PCI bus to the processor main memory, and then
sent to another computer outside the oven through a LAN card.
The test was performed at average trigger frequency 10 Hz.

No error occurred in the test.
The complete system worked correctly.

\section{Summary}

 We have developed a widely usable high-density DAQ system equipped
with a PMC bus as the internal bus.
It has data buffers just behind the digitizers to reduce DAQ dead-time.
On-board data are reduced by a common PC architecture. 
The system is modular and so the processor and the data transfer interface
are upgradeable.
The system is flexible and scalable and may be applied to a wide range of
experiments because of its modular structure and data link
interface for data transfer.

The performance of data transfer from FIFOs to processor
 main memory is 125 MB/sec without errors.

We are developing a refined version of the system
 with digitizing front-end daughter cards,
and a trigger handling and distribution system.

\begin{acknowledgments}
This work was supported in Japan by a Grant-in-Aid from the Ministry
of Education, Science, Sports and Culture.
We would like to express their gratitude to all the members of
Belle collaboration and all members involved in the J-PARC Project.
The authors thank S. Yamada, S. Iwata, F. Takasaki, M. Kobayashi
and K.Nakamura.

\end{acknowledgments}


\end{document}